\documentclass[journal,A4]{IEEEtran}

\usepackage{cite}
\usepackage{booktabs}
\usepackage{graphicx}
\usepackage{amssymb,amsmath} 
\usepackage{multirow}
\usepackage{url}
\usepackage{float}

\usepackage{stfloats}

\makeatletter
\let\MYcaption\@makecaption
\makeatother

\usepackage[font=footnotesize]{subcaption}

\makeatletter
\let\@makecaption\MYcaption
\makeatother

\begin{document}

\bstctlcite{IEEEexample:BSTcontrol}

\title{Validation of  e$^+$e$^ -$ Pair Production Total Cross Sections for Monte Carlo Particle Transport}

\author{Tulio Basaglia, 
	Marcia Begalli, 
	Chansoo Choi, 
	Min Cheol Han, 
	Gabriela Hoff, 
	Chan Hyeong Kim, \\
	Maria Grazia Pia 
	and Paolo Saracco
\thanks{Manuscript received 1 January 2099.}
\thanks{T. Basaglia is with CERN, CH-1211, Geneva, Switzerland (e-mail: Tullio.Basaglia@cern.ch).}
\thanks{M. Begalli is with State University of Rio de Janeiro, Brazil.}
\thanks{C. Choi and C. H. Kim are with the Department
of Nuclear Engineering, Hanyang University, Seoul 04763, Korea (e-mail:
cchoi91@hanyang.ac.kr, chkim@hanyang.ac.kr).}
\thanks{M. C. Han is with the Department of Radiation Oncology, Yonsei University College of Medicine, Seoul 03722, Korea (email: mchan@yuhs.ac)}
\thanks{G. Hoff, M. G. Pia and Paolo Saracco are with INFN Sezione di Genova, Via Dodecaneso 33, I-16146 Genova, Italy 
	(email: Gabriela.Hoff@ge.infn.it, MariaGrazia.Pia@ge.infn.it, Paolo.Saracco@ge.infn.it).}
}

\maketitle


\begin{abstract}
Several sources of total cross sections for e$^+$e$^-$ pair production by
photon interaction with matter, used by major Monte Carlo codes for particle
transport, are quantitatively evaluated with respect to experimental
measurements collected from the literature.
They include data libraries deriving from theoretical calculations and analytical formulae 
representing empirical fits to tabulations.
Although most of these sources of cross sections are based on one theoretical
reference, the statistical data analysis documented in the paper highlights
differences in their compatibility with experiment.
The cross sections collected in the 1997 version of the Evaluated Photon Data
Library (EPDL) exhibit the lowest incompatibility with experiment; the paper
details the validation results for all the examined cross section sources.
The scarcity of experimental data does not allow the validation process to
discern pair production in the field of the nucleus and in the field of
atomic electrons; nevertheless, the paper documents relevant clues regarding the
contribution of the latter to reproduce experimental measurements.
\end{abstract}


\section{Introduction}
\label{sec_intro}

\PARstart{P}{hoton} interactions with matter involve several atomic physics
processes; in the region of photon energy above a few MeV, e$^+$e$^-$ pair
production is the dominant one \cite{carron_book}. 
Hence, the simulation of this process is important in various experimental
domains that involve high energy photons, such as fundamental physics
investigations in particle and nuclear physics experiments,
medical physics, radiation dosimetry and shielding.

The occurrence of this process in the course of photon transport is determined
by its total cross section. 
The validation test documented in this paper evaluates the capability of
widely used Monte Carlo systems for particle transport to model e$^+$e$^-$ pair
production total cross section consistently with experiment;
its results identify the state of the art among the calculation methods adopted
by these codes.
To this end, it employs methods of statistical inference to evaluate cross section
calculations quantitatively and objectively with respect to a collection of
experimental measurements.

To the best of our knowledge, this paper reports the first large-scale
validation test of e$^+$e$^-$ pair production total cross sections used in major
particle transport systems.
Previous papers \cite{tns_nist, cirrone2010}, limited to photon cross sections
used by Geant4 \cite{g4nim, g4tns, g4nim2}, document their verification with
respect to theoretical calculations, while experimental publications, such as
those listed in the bibliography, at most report qualitative appraisals of a
small set of measurements with respect to theory.

This extensive investigation provides guidance to the developers and the users
of Monte Carlo codes to address the requirements of physics accuracy pertinent
to different experimental scenarios.



\section{Physics overview}

Photon conversion has been the object of theoretical and experimental interest
for several decades; only a brief overview is summarized here to facilitate the
comprehension of the results documented in this paper.

\subsection{Particle transport environment}

The cross section calculation methods examined in this paper reflect the assumptions and
approximations adopted in the general-purpose Monte Carlo transport systems
used in experimental particle and nuclear physics, and related  fields.
They deal with interactions with free atoms; they neglect the effects of the
environment of the target atom and multi-photon effects associated with incident
beams.
They do not take into account either any internal degrees of freedom of the
initial or final atom, such as its orientation; therefore they do not describe
photon interactions in molecules and oriented solids.

The discussion of calculation methods and their validation is limited in this
paper to the energy range where total cross section measurements could be
retrieved from the literature.

\subsection{Theoretical calculations}


The total photon cross section gathers contributions from various independent
processes: 
at lower photon energies, it is dominated by the photoelectric effect, where a
photon ejects an electron from the target atom; then, Compton scattering becomes
relevant, where a photon is scattered by an almost free electron; and then, an
e$^+$e$^-$ pair can be created in an external electromagnetic field if the
energy of the photon becomes higher than the couple rest mass.
At even higher energies, other inelastic processes may
occur, such as  photonuclear reactions [$(\gamma,n)$ or $(\gamma,p)$ above
approximately
10~MeV],  photomeson production and conversion to couples of
heavier particle-antiparticle pairs. 


Electron-positron pair production can happen thanks to the presence of an
external electromagnetic field that permits a virtual particle-antiparticle
couple to become real, as only in such a case energy-momentum conservation can be satisfied.
The electromagnetic field can be provided by the atomic nucleus of the target
atom or by the electrons surrounding it.
The latter case is known as triplet pair production;
the ``triplet'' naming 
stems from the peculiar signature of this reaction in a cloud
chamber, caused by the ejection of the atomic electron as well.
The two processes  occur  with different intensities.

This phenomenon was theoretically predicted by Dirac in 1928 \cite{dirac1928}
and observed by Anderson in 1933 \cite{anderson1933}, who was later credited
with the Nobel prize; positron tracks were present in earlier
photographs, but experimental limitations did not permit reaching definite 
conclusions \cite{bazilevskaya2014}.
More detailed theory was developed by Bethe and Heitler \cite{bethe_heitler};
refinements to the calculations have been pursued for the next decades
\cite{hubbell_pair_2006}.

The threshold  energy for a conversion process depends on
the mass $M$ of the particle that  is initially at rest and is different for
singlet and triplet pair production:
\begin{equation}
\begin{split}
 T_{lab}^s\simeq 2 m_e c^2\quad{\rm if} \:  M>>m_e \\
T_{lab}^t\simeq 4 m_e c^2\quad{\rm if} \: M=m_e,
\end{split}
\end{equation}
corresponding to approximately 1.022~MeV and 2.044~MeV, respectively.
The cross section for singlet pair production varies with atomic number as ${\sigma_s\sim Z^2}$,
as $\sigma_t\sim Z$ for triplet.

The processes of singlet and triplet pair production are also characterized
by a different final state, because triplet production necessarily
involves excitation and ionization of the target atom, while singlet pair production
involves only interaction with the atomic nucleus.
The ionization of the target atom yields a small shift in the threshold energy
\cite{hubbell_1980} and two identical electrons in the final state, which should 
then be anti-symmetrized in the calculations.

Several theoretical calculations of pair production cross sections have been 
documented in the literature; a comprehensive review can be found in \cite{hubbell_pair_2006}.
A systematic evaluation of the computational approaches is discussed in a
landmark paper by Hubbell, Gimm and \O{}verb\o{} \cite{hubbell_1980}.
The calculation strategy of \cite{hubbell_1980} starts from the Born
approximation in an unscreened field, on top of which Coulomb, screening and
radiative corrections are successively implemented.
Coulomb correction amounts to taking care of successive
iteration of Coulomb interaction: the importance of such terms stems from the
fact that the expansion parameter is in $\alpha Z$ for pair production 
(where $\alpha$ is the fine structure constant); it is not important, and
is usually omitted, in triplet cross section calculations \cite{maximon1981, gimm_1978}.
The treatment of screening takes care of the presence of other charges.
Screening corrections vary significantly with energy, but  they can
be evaluated with good accuracy.
Finally, lowest order radiative corrections amount to the emission and reabsorption of
virtual photons.

From a theoretical point of view, many effects should be taken into account
in the calculation of triplet production cross sections:
the atomic binding of the target electron, 
the screening by other electrons and by the field of the nucleus, 
the retardation due the recoil of the target electron,  
the interaction of the incident photon with the atomic electron via virtual 
Compton scattering and production of virtual electron-positron pairs, 
the exchange terms due to the indistinguishability of the two electrons and 
radiative corrections. 
The calculation are of considerable complexity; since a comprehensive
computation taking into account all these effects is lacking, the cross sections in
\cite{hubbell_1980} derive from combining a variety of theoretical approaches
and numerical evaluations.



\section{Cross sections in particle transport}
\label{sec_mc}


The calculation of cross sections based on state-of-the-art theoretical methods
in the course of transport would be a prohibitive burden for the computational
performance of simulation applications.
Therefore, Monte Carlo codes calculate the cross sections needed for particle
transport either by interpolation of data libraries or through simple analytical
formulae, which may derive from empirical parameterizations, from simplified
models or from fits to tabulated data.
Most of these codes account for both pair and triplet production cross sections
to determine the occurrence of a photon interaction, although they usually
simplify the generation of the final state as if pair production would always
occur in the nuclear field.

Tabulations of pair production total cross sections, 
which incorporate
the body of knowledge of theoretical approaches at the time of their publication,
were produced by Hubbell, Gimm and \O{}verb\o{} \cite{hubbell_1980} in 1980;
they still represent the most authoritative reference for Monte Carlo transport codes.
They report the contributions of pair production in the field of the nucleus and
in the field of the atomic electrons for the elements with atomic number up to
100, and for energies from the threshold (corresponding to twice the electron
mass) to 100 GeV.
These calculations have been incorporated in the Evaluated Photon Data Library
(EPDL) \cite{epdl89, epdl97}, in the XCOM database \cite{xcom} of the National
Institute of Standards and Technology) (NIST) and in the PHOTX
\cite{trubey_1989} data library, which in turn was developed for inclusion in
the ENDF/B-VI \cite{endfb6} physics  data library.
Although these data libraries share the same physical foundation of
\cite{hubbell_1980}, they  may exhibit different features, such as the
number of data values, the photon energies and the number of significant digits
in the tabulations.

EPDL is extensively used in Monte Carlo simulation.
Several well-known particle transport codes, such as
EGSnrc \cite{egsnrc}, 
FLUKA \cite{fluka1, fluka2}, 
Geant4 \cite{g4nim,g4tns,g4nim2}, 
ITS \cite{its6}, 
MCNP \cite{mcnp6},  
PHITS \cite{phits} and
Serpent \cite{kaltiaisenaho2020},
base the simulation of e$^+$e$^-$ pair production on the cross sections
tabulated in EPDL.

Most particle transport codes use the EPDL version released in 1997, also known
as EPDL97 \cite{epdl97}, which was included in ENDF/B-VI.8 and in the following
releases of the ENDF/B-VII series \cite{chadwick2006, chadwick2011}.
An EPDL version released by IAEA (International Atomic Energy Agency) in 2014
appears to be identical to the 1997 version, apart from the format of numbers in
scientific notation.

New versions of EPDL were released in 2018: one was included in the ENDF/B-VIII.0
\cite{endfb8} data library, encoded in the ENDF-6 format, and others were distributed by
IAEA within EPICS 2017, in ENDF-6 and ENDL format, respectively.
The new releases raised several issues, documented in \cite{tns_endfb8}; those
included in EPICS 2017 appear to have undergone some modifications with respect
to the originally released content, while keeping the same ``EPICS 2017''
identification.
One of the issues highlighted in \cite{tns_endfb8} is, indeed, the lack of
proper version control in the IAEA distribution of EPDL and of the associated
electron and atomic data libraries, which prevent the unambiguous identification
of different data downloaded from this distribution source as ``EPICS 2017''.
According to the associated documentation \cite{cullenepdl_2018}, the EPDL 
data released in 2018 can be linearly interpolated, while logarithmic interpolation
was recommended for EPDL97.


In addition to EPDL, other data libraries are used in major particle transport
codes to calculate pair and triplet production cross sections: EGS5 \cite{egs5}
uses the PHOTX compilation, while EGSnrc provides the option of using the
compilation by Storm and Israel \cite{storm_1970}, XCOM or the 1997 version of
EPDL.

The Penelope \cite{penelope2018} Monte Carlo system encompasses a compilation of
pair and triplet production cross sections, which are released along with the
source code.
The approach adopted in the calculations, detailed in the code documentation,
appears to be the same as in \cite{hubbell_1980}.

Other simulation codes calculate cross sections according to analytical
formulae rather than through interpolation of tabulated data. 
This is the case for Geant4 Standard electromagnetic package, which adopts a 
similar solution to its predecessor GEANT 3 \cite{g3}; for both codes, the 
parameterizations of pair production cross sections are based on fits to the
data tabulated by  Hubbell, Gimm and \O{}verb\o{} in \cite{hubbell_1980}.
An ultra-relativistic model, also implemented in Geant4, computes pair
production cross sections using direct integration of differential cross
sections \cite{g4physrefmanual}; it is not considered in the validation test
reported in this paper due to lack of experimental measurements in the energy
range recommended for its use, above 80~GeV.

To the best of our knowledge, systematic and quantitative validation of the above
mentioned compilations of pair  production cross sections is not yet
documented in the literature.
\section{Strategy of this study}
\label{sec:strategy}

\begin{table*}[htbp]
  \centering
  \caption{Cross section sources}
    \begin{tabular}{llccc}
    \toprule
    Identifier & Description & Type  & Content & Energy \\
    \midrule
    EPDL     & EPDL, 1997 version & tabulation & singlet and triplet pair production cross sections & 1.022 MeV - 100 GeV \\
    ENDFB  & ENDF/B-VIII.0 & tabulation & singlet and triplet pair production cross sections & 1.022 MeV - 100 GeV \\
    EPICSF   & EPICS 2017, ENDF format & tabulation & singlet and triplet pair production cross sections & 1.022 MeV - 100 GeV \\
    EPICSL   & EPICS 2017, ENDF format & tabulation & singlet and triplet pair production cross sections & 1.022 MeV - 100 GeV \\
    Pen18  & Penelope 2018 & tabulation & singlet and triplet pair production cross sections & 1.022 MeV - 100 GeV \\
    PHOTX & PHOTX & tabulation & singlet and triplet pair production cross sections & 1.022 MeV - 100 GeV \\
    Storm & Storm and Israel & tabulation & singlet and triplet pair production cross sections & 1.022 MeV - 100 MeV \\
    XCOM  & XCOM, default energies & tabulation & singlet and triplet pair production cross sections & 1.022 MeV - 100 GeV \\
    \midrule
    G3    & GEANT 3 & analytical & sum of singlet and triplet pair production cross sections & 1.022 MeV - 100 GeV \\
    Std   & Geant4 Standard & analytical & sum of singlet and triplet pair production cross sections & 1.022 MeV - 100 GeV \\
    \bottomrule
    \end{tabular}%
  \label{tab:models}%
\end{table*}%

\setlength{\tabcolsep}{6pt}
\begin{table*}[tbhp]
\begin{center}
\caption{Summary of the experimental cross section data used in the validation process}
\label{tab_exp1}
\begin{tabular}{rlccl}
\toprule
\multicolumn{2}{c}{Element} & Energy range & Sample  & References \\ 
$Z$  &  Symbol               & (MeV)      & size         &            \\ 
\midrule
1 & H & 330-8700 & 22 & \cite{eickmeyer1980}, \cite{fujii1976}, \cite{meyer1970} \\
2 & He & 2.62 & 1 & \cite{dayton1953} \\
3 & Li & 17.6 & 1 & \cite{walker1949} \\
6 & C & 1.173-8700 & 14 & \cite{dayton1953}, \cite{eickmeyer1980}, \cite{knasel1968}, \cite{walker1949}, \cite{west1956} \\
13 & Al & 1.119-8700 & 12 & \cite{dayton1953}, \cite{eickmeyer1980}, \cite{girard1978}, \cite{hahn1952}, \cite{knasel1968}, \cite{titus1966}, \cite{walker1949} \\
22 & Ti & 1.173-2.754 & 5 & \cite{henry1972} \\
26 & Fe & 1.119-2.615 & 6 & \cite{avignone1982}, \cite{girard1978}, \cite{hahn1952}, \cite{khalil1982} \\
28 & Ni & 1.779 & 1 & \cite{hahn1952} \\
29 & Cu & 1.119-8700 & 21 & \cite{avignone1982}, \cite{dayton1953}, \cite{eickmeyer1980}, \cite{girard1978}, \cite{hahn1952}, \cite{henry1972}, \cite{khalil1982}, \cite{knasel1968}, \cite{rao1963}, \cite{schmid1954}, \cite{walker1949} \\
32 & Ge & 1.0404-11.775 & 170 & \cite{debraeckeleer1992}, \cite{coquette1977}, \cite{coquette1978}, \cite{coquette1979}, \cite{coquette1980}, \cite{enyo1980}, \cite{frahm2009}, \cite{jentschel2011}, \cite{mittal1989}, \cite{sharma1985}, \cite{yamazaki1965} \\
34 & Se & 1.779 & 1 & \cite{hahn1952} \\
40 & Zr & 1.119 & 1 & \cite{rao1963} \\
42 & Mo & 1.12-2.754 & 7 & \cite{hahn1952}, \cite{henry1972} \\
45 & Rh & 1.119 & 1 & \cite{rao1963} \\
47 & Ag & 1000 & 1 & \cite{dauvergne2003} \\
48 & Cd & 1.12-2.754 & 9 & \cite{hahn1952}, \cite{henry1972} \\
50 & Sn & 1.115-17.6 & 22 & \cite{avignone1974}, \cite{avignone1974a}, \cite{avignone1982}, \cite{bose1982}, \cite{dayton1953}, \cite{garritson1968}, \cite{girard1978}, \cite{hahn1952}, \cite{khalil1982}, \cite{rao1963}, \cite{schmid1954}, \cite{titus1966}, \cite{walker1949} \\
53 & I & 1.077-2.754 & 10 & \cite{huck1964}, \cite{west1956} \\
58 & Ce & 1.779 & 1 & \cite{hahn1952} \\
73 & Ta & 1.119-2.614 & 2 & \cite{rao1963}, \cite{titus1966} \\
74 & W & 1.119-2.754 & 7 & \cite{hahn1952}, \cite{henry1972} \\
78 & Pt & 1.119 & 1 & \cite{rao1963} \\
79 & Au & 1.115-1000 & 4 & \cite{bose1982}, \cite{dauvergne2003}, \cite{rao1963}, \cite{titus1966} \\
82 & Pb & 1.077-1200 & 33 & \cite{avignone1974a}, \cite{avignone1982}, \cite{avignone1985}, \cite{dayton1953}, \cite{garritson1968}, \cite{girard1978}, \cite{hahn1952}, \cite{henry1972}, \cite{jenkins1956}, \cite{khalil1982}, \cite{knasel1968}, \cite{moore1956}, \cite{rao1963}, \cite{schmid1954}, \cite{shkolnik1957}, \cite{standil1958}, \cite{walker1949} \\
83 & Bi & 1.275-2.615 & 3 & \cite{hahn1952} \\
90 & Th & 1.115 & 1 & \cite{bose1982} \\
92 & U & 1.115-8700 & 10 & \cite{avignone1982}, \cite{bose1982}, \cite{eickmeyer1980}, \cite{girard1979}, \cite{hahn1952}, \cite{khalil1982} \\

\bottomrule
\end{tabular}
\end{center}
\end{table*}
\setlength{\tabcolsep}{6pt}

Consistent with the pertinent Standard \cite{ieeestd1012}, the validation of
cross sections  intended for use in particle transport
has a pragmatic nature: Monte Carlo codes are concerned with identifying the
state of the art among practically usable modeling methods -- tabulations or
simple analytical formulations -- on the basis of the available body of knowledge,
i.e. of existing experimental data.
The extent of the validation tests and the depth of their physics investigation are
determined by the amount and the characteristics of the available experimental
measurements; they are meaningful if they have sufficient power to achieve
significant conclusions.

The validation process of pair production cross sections addresses two issues:
appraising the compatibility with experiment of the calculation methods used in
major Monte Carlo codes for particle transport, and identifying the state of the
art for simulation applications among them.

For these purposes, the data libraries and analytical formulations representing
the approaches adopted in major Monte Carlo transport codes have been singled
out; cross section calculations based on them have been implemented in a
consistent software design, which minimizes external dependencies to ensure the
unbiased appraisal of the intrinsic capabilities of the methods subject to evaluation.


A set of experimental data has been collected from the literature for the
validation of the calculation methods.
The majority of the experimental measurements do not distinguish the production
of e$^+$e$^-$ pairs deriving from interactions in the field of the nucleus or in
the field of the atomic electrons; therefore, the validation process concerns
the calculations representing the sum of the cross sections pertaining to both
processes.
Nevertheless, the analysis highlights the critical role played by the 
calculations of triplet production.

The use of statistical inference methods in the validation process ensures
objective and quantitative conclusions.
Goodness-of-fit tests compare calculated and experimental data distributions to
evaluate the incompatibility with experiment of cross section data libraries and
analytical formulations; techniques of categorical data analysis are applied
to ascertain whether the various methods' cross section calculations exhibit
significant differences in incompatibility with experiment.

The validation process is supported by investigations concerning the robustness
of the results and by assessments of the power of the tests.

The limited amount of available experimental data prevents the stratification of
the analysis in terms of atomic number, photon energy or other specific
features: the power of the tests considerably drops when one partitions the
small data sample, thus hindering the achievement of meaningful results over
strata.
This limitation prevents the evaluation of possible effects related to the
inhomogeneity of the existing experimental sample, which is determined by the
experimental activity of the past decades and could only be mitigated by
performing further measurements in the future;
nevertheless, it does not affect the conceptual cogency of the feasible tests in
the peculiar context of Monte Carlo codes, where the concern is using
state-of-the-art of pair production cross sections in particle transport,
compatible with existing knowledge.



In the following sections, the type of occurrence of the pair production
process, i.e. production in the field of the nucleus (or singlet production), and
production in the field of atomic electrons (or triplet production), is
explicitly specified when the distinction is relevant to the context.



\subsection{Cross section calculations}

The cross section sources subject to test are summarized in Table
\ref{tab:models}.
Data libraries provide distinct tabulations of pair production cross sections in
the field of the nucleus and of the atomic electrons, while the analytical cross section
formulations considered in the validation process calculate the sum of the two
contributions.

All the cross section calculation methods involved in the validation process
have been implemented in a stand-alone test system, adopting a policy-based
class design \cite{alexandrescu}.
This software design supports the provision of a wide variety of physics
modeling approaches without imposing the constraint of inheritance from a
predefined interface.
A single policy class calculates cross sections based on data libraries;
alternative data tabulations are managed through the file system.
Dedicated policy classes implement cross section calculations based on
analytical formulae.

The correctness of the implementation of cross section calculations has been
verified to ensure the software reproduces the physical features of each
calculation method consistently.

In the validation process, the cross sections subject to test  are
calculated in the same settings (photon energy and target element) as the
experimental data they are compared to.


\subsection{Experimental data}
\label{sec_exp}

The experimental data sample, derived from an extensive survey of the literature
\cite{avignone1974, avignone1974a, 
avignone1982, 
avignone1985, 
bose1982, coquette1977, coquette1978, coquette1979, coquette1980,
dauvergne2003, dayton1953, debraeckeleer1992,
eickmeyer1980, enyo1980,
frahm2009, fujii1976,
garritson1968, girard1978, girard1979,
hahn1952, henry1972, huck1964,
jenkins1956, jentschel2011,
khalil1982, knasel1968,
meyer1970, mittal1989, moore1956,
rao1963, 
schmid1954, sharma1985, shkolnik1957, standil1958,
titus1966,  walker1949, west1956, yamazaki1965},
consists of 367 measurements, which concern target elements from hydrogen to
uranium and span the energy range from 1.04 MeV to 8.7 GeV.
An overview of this data sample is summarized in Table \ref{tab_exp1}.
Despite the effort invested in the search of relevant data in the literature,
this experimental sample is much smaller than the pools of measurements used in
similar validation tests of photon interaction cross sections
\cite{tns_rayleigh, tns_photoel, tns_photoel2}.

The raw experimental sample collected from the literature was further examined
to ascertain the usability of the data.
This assessment is particularly delicate due to the scarcity of measurements, as
further reduction of the sample size could compromise the power of the
statistical tests, i.e. their capability to produce significant results.

Precise knowledge of the primary photon energy is necessary in the 
validation process to ensure that the cross sections subject to test are calculated
under the same conditions as in the experimental setup.
The experimental publications involving radioactive sources often report
inconsistent values for the energy of the emitted photons: the imprecision of
these data reflects the body of knowledge of radioactive decays at the time when
the experiments were performed.
The energies of photons originating from radioactive sources were updated in the
validation test process to conform to recent authoritative standards
\cite{iaea2007update}, thus ensuring consistency and correctness of 
the cross section calculations involved in the test.

Some experimental cross sections are published only in graphical form;
numerical values were digitized from the plots by means of the Engauge Digitizer
\cite{engauge} software.
The reliability of the digitized cross section values is hindered by the
difficulty of appraising the experimental points and their error bars in plots that
may span several orders of magnitude in logarithmic scale.
The error introduced by the digitization process was  estimated by digitizing a
few plotted data, whose numerical value is explicitly reported in the related
publications; nevertheless, due to the variable quality of the 
plots in the collected experimental references, this estimate can only 
be considered an approximate indication.
The effects on the validation tests introduced by using digitized experimental data
were evaluated in the analysis and are reported in Section \ref{sec:chi2robust}.

Several references report semi-empirical pair production cross sections,
deriving from measurements of total mass attenuation coefficients: they were
estimated in the original papers by subtracting the contributions of other processes (Compton
scattering and photoelectric absorption) from the measured observables.
Since these data are not pristine experimental measurements, they were discarded
from the validation process.

Some experimental measurements are reported in the associated publications
relative to theoretical calculations rather than as absolute values, usually
with the intent to highlight consistency or deviations with respect to theory.
A number of them can be converted into their pristine absolute values, based on
the information available in the original papers; for others, lacking adequate details,
one can only formulate educated guesses on the theoretical values,
relative to which the published measurements are expressed.
These experimental references are liable to introduce systematic effects in the
validation process; their treatment in the analysis is discussed in Section
\ref{sec:chi2pair} and the effects of their inclusion in the statistical analysis
are reported in Section \ref{sec:chi2robust}.

A special case concerns a few cross section measurements that are reported
relative to a reference value (at a given energy and for a given target
element), whose absolute value is fixed  by a theoretical  calculation or
by another experiment.
Since there is consensus that some theoretical calculations used in the past for this purpose
are now obsolete and have been superseded by more authoritative theoretical
approaches, experimental cross sections normalized to such theoretical values
are prone to introduce systematic effects. 
Therefore, the reference values were renormalized in the validation process with
respect to the more recent calculations based on EPDL, and the number of degrees
of freedom in the goodness-of-fit tests was adjusted accordingly to calculate
the p-value of the tests correctly.


Correct estimate of experimental errors is a concern in the validation process,
since an unrealistic estimation may lead to incorrect conclusions in the
$\chi^2$ test regarding the rejection of the null hypothesis of compatibility
between calculations and experiment.
Some experimental publications do not document the experimental uncertainties of
the cross section measurements they report; estimated errors were introduced in
the validation process, deriving from an educated guess based on the typical
uncertainties of similar measurements performed in similar conditions, using
similar experimental techniques.
When such educated guesses were not possible, the experimental data lacking
uncertainties were discarded from the validation process.
Other experimental references explicitly state that the results they report only
account for statistical uncertainties, thus hinting at underestimated
uncertainties affecting the outcome of the $\chi^2$ test.

\begin{figure}[tbp]
\centerline{\includegraphics[angle=0,width=8.5cm]{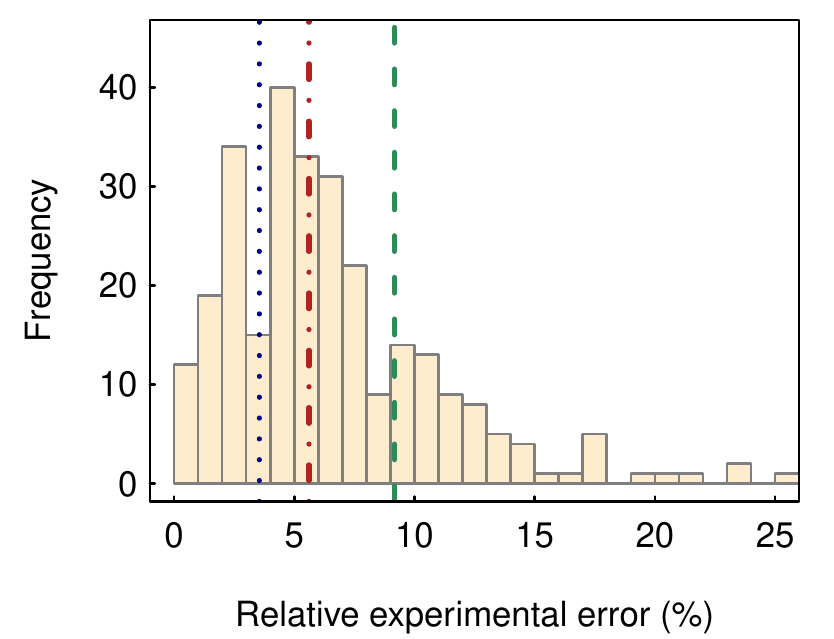}}
\caption{Distribution of the relative  error of the experimental measurements:
the red dot-dashed line denotes the median value of the distribution; the blue
dotted and the green dashed lines indicate the first and last quantile in the
distribution, respectively the lowest and the highest 25\% portions of the distribution.}
\label{fig:experr}
\end{figure}

The experimental data sample encompasses measurements of varying precision;
Fig. \ref{fig:experr} shows the  distribution of the relative errors of the 
experimental data, derived from the respective publications.
One cannot discern any manifest association between the relative error reported
in the experimental papers and the date of the experiments: that is, more recent
measurements are not necessarily more precise than older ones.
The effect of measurements with varying precision on the validation process 
is discussed in Section \ref{sec:chi2robust}.

Discrepancies are evident in some of the experimental data;
systematic effects are likely present in some cases, where 
sequences of positive or negative differences between
data samples originating from different experimental groups hint at 
incompatibility with randomness.
Some of these discrepancies were already highlighted and discussed in the
experimental publications, e.g. in \cite{debraeckeleer1992, jentschel2011}.
Due to the limited availability of experimental measurements and the scarcity of
documentation in some experimental publications, it is not always possible to
ascertain which data sets exhibiting large discrepancies could be affected by
systematic effects; therefore, it is unavoidable that the experimental data
sample used in the validation process may be polluted by unreliable measurements.

Quality criteria are applied to the experimental data involved in the validation process
to mitigate the risks associated with questionable experimental sources, as 
is documented in Section \ref{sec_resultspair}.
The evaluation of the power of the tests, reported in Section \ref{sec:cont_pair},
also supports the robustness of the results.

%
%




\subsection{Data analysis method}
\label{sec:analysis}


The validation process applies the methodology described in \cite{tns_photoel}:
first, cross sections calculated by each model are compared with experimental
data to determine their incompatibility with experiment;
in the following stage,
categorical data analysis is performed to identify significant differences in 
compatibility with experiment among the various calculation methods.
Further details concerning the methodology 
can be found in  
\cite{tns_sandia1,
tns_photoel, 
tns_photoel2,
tns_rayleigh,
tns_bebshell,
tns_ebscatter3,
tns_ebscatter,
tns_sandia2,
tns_beb,
tns_pcross,
tns_pixe,
tns_trans_prob}.
The level of significance of the tests  is 0.01, unless stated otherwise.


The first stage of the analysis is articulated through a series of $\chi^{2}$
tests \cite{bock} over test cases identified consistently with the available 
experimental data distributions.
The test cases reflect how the data are reported by the experiments: either as
measurements concerning specific elements as a function of energy or as measurements at
fixed energies as a function of the atomic number of the target elements.
The cross sections subject to validation are calculated in the same settings
(photon energy and target element) as the experimental data.

The null hypothesis in the $\chi^{2}$ test is defined as the
calculated and experimental data distributions being equivalent.
For convenience, the ``efficiency'' of a cross section model is defined as the
fraction of test cases in which the $\chi^2$ test does not reject the null
hypothesis at the selected level of significance.

It is worthwhile to recall that the $\chi^{2}$ test takes into account the
experimental uncertainties explicitly in the calculation of the test statistic;
therefore, its outcome is sensitive to incorrect estimates of the experimental
errors.

In the second stage of the analysis the results of the $\chi^2$ test are
summarized in $2\times2$ contingency tables, which report the number of test
cases classified as ``fail'' or ``pass'', according to whether the hypothesis of
compatibility of experimental and calculated cross sections is or is not rejected.
The null hypothesis in the test of a contingency table assumes equivalent
compatibility with experiment of the cross section calculation methods subject
to test.
Exact tests (Fisher \cite{fisher1922}, Boschloo \cite{boschloo}, Z-pooled
\cite{suissa} and Barnard \cite{barnard} in the CSM approximation
\cite{barnard_csm}) are used in the analysis of contingency tables; Pearson's
$\chi^2$ test \cite{pearson} is also used, when the number of entries in the
cells of the table justifies its applicability (i.e., it is greater than 5 \cite{lyonsbook}).
The  variety of tests applied to contingency tables mitigates the risk of introducing
systematic effects in the analysis, which could be related to features of the mathematical
formulation of the tests.


The analysis reported in this paper used the R software system \cite{R}, version 4.1.0.

\section{Results}
\label{sec_resultspair}

\begin{figure}[p]
\centerline{\includegraphics[angle=0,width=8.5cm]{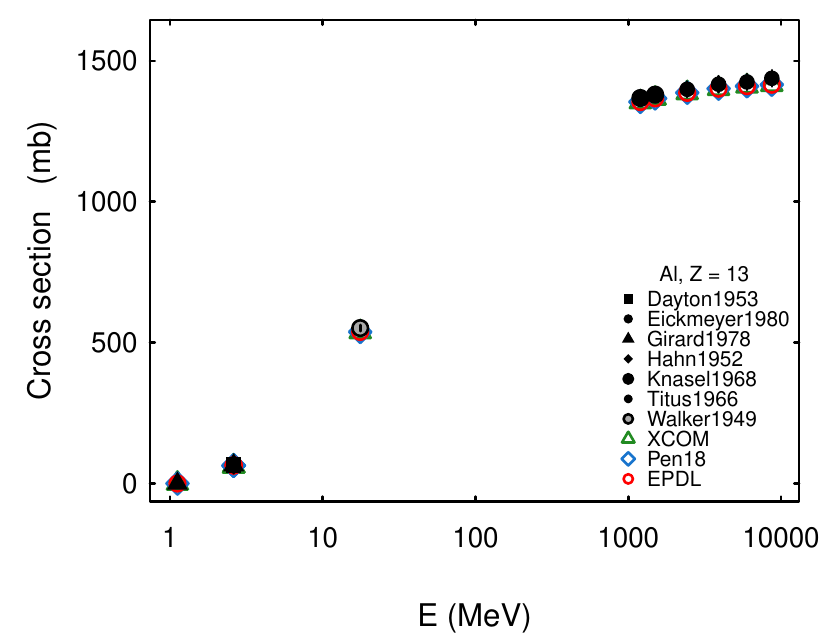}}
\caption{Total $e^+e^-$ pair production cross sections  for aluminium as a function of photon energy.}
\label{fig_tot13}
\end{figure}
\begin{figure}[p]
\centerline{\includegraphics[angle=0,width=8.5cm]{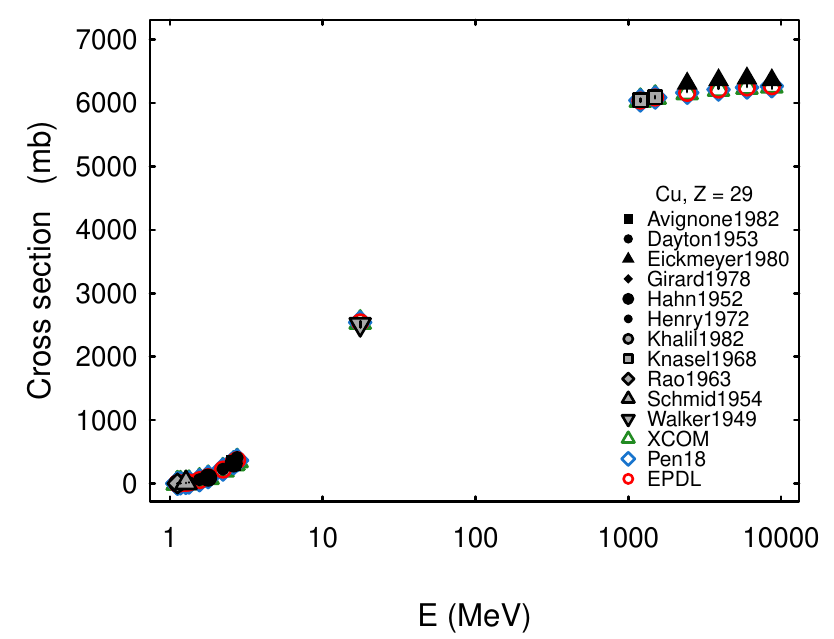}}
\caption{Total $e^+e^-$ pair production cross sections  for copper as a function of photon energy.}
\label{fig_tot29}
\end{figure}
\begin{figure}[p]
\centerline{\includegraphics[angle=0,width=8.5cm]{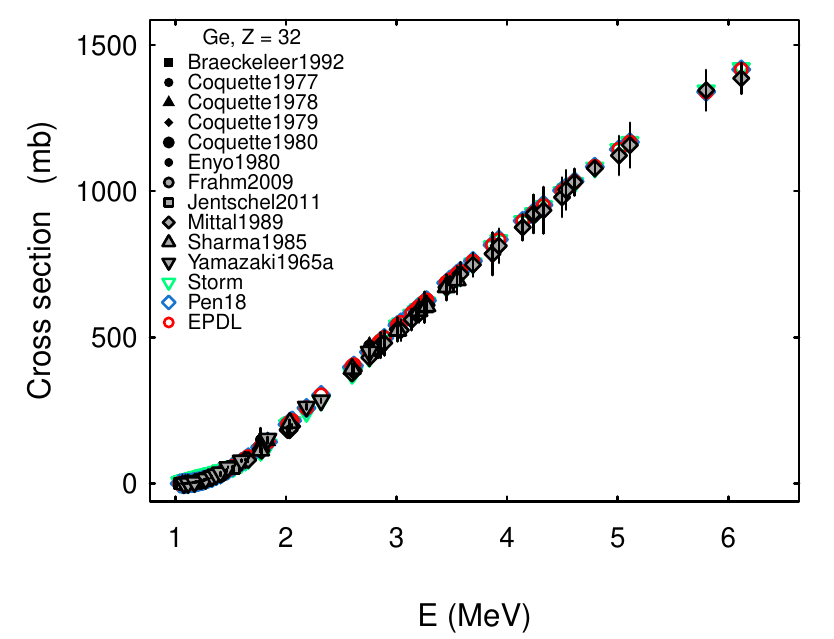}}
\caption{Total $e^+e^-$ pair production cross sections  for germanium as a function of photon energy.}
\label{fig_tot32}
\end{figure}
\begin{figure}[p]
\centerline{\includegraphics[angle=0,width=8.5cm]{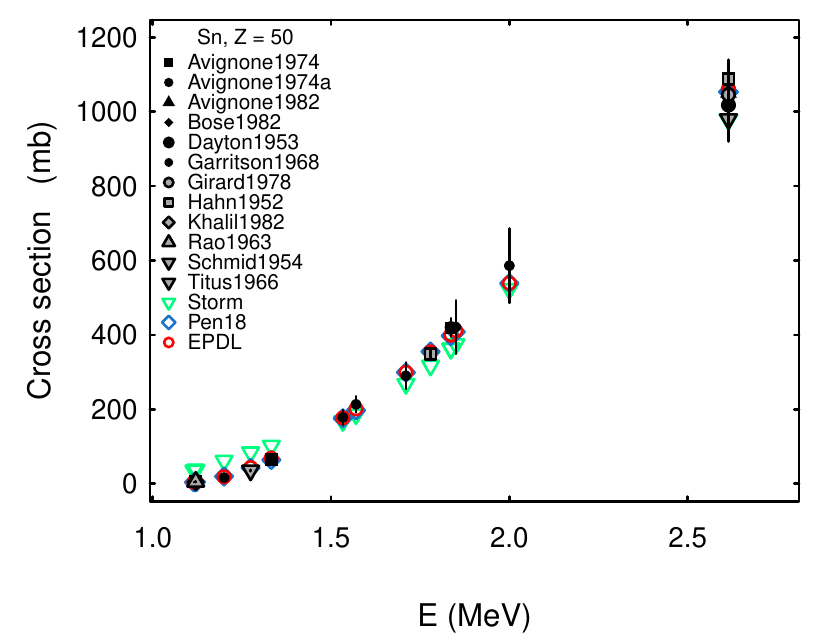}}
\caption{Total $e^+e^-$ pair production cross sections  for tin as a function of photon energy.}
\label{fig_tot50}
\end{figure}
\begin{figure}[p]
\centerline{\includegraphics[angle=0,width=8.5cm]{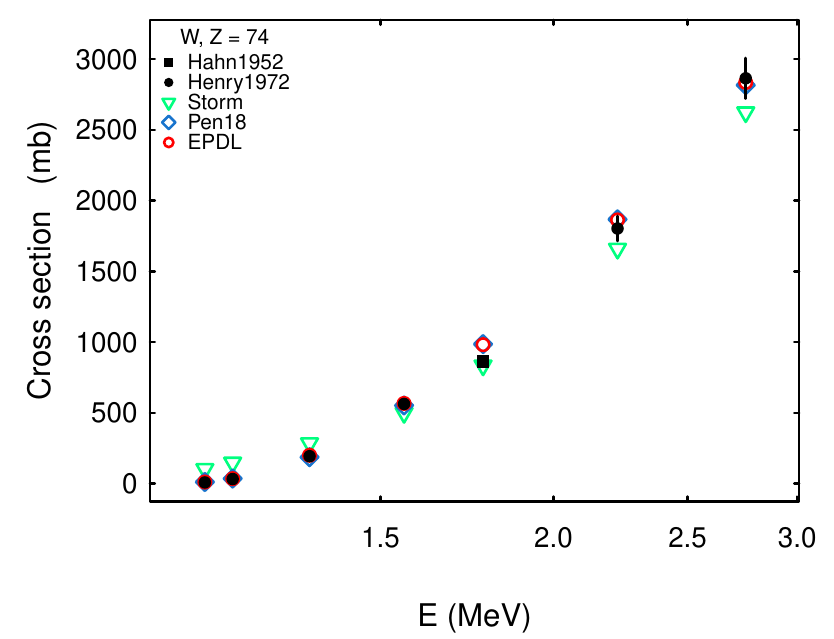}}
\caption{Total $e^+e^-$ pair production cross sections  for tungsten as a function of photon energy.}
\label{fig_tot74}
\end{figure}
\begin{figure}[p]
\centerline{\includegraphics[angle=0,width=8.5cm]{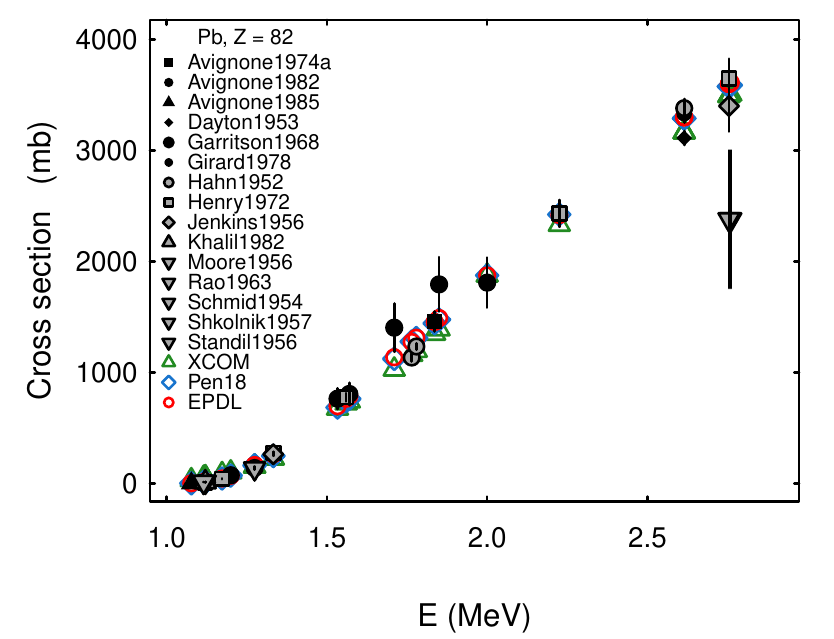}}
\caption{Total $e^+e^-$ pair production cross sections  for lead as a function of photon energy.}
\label{fig_tot82}
\end{figure}
\begin{figure}[t]
\centerline{\includegraphics[angle=0,width=8.5cm]{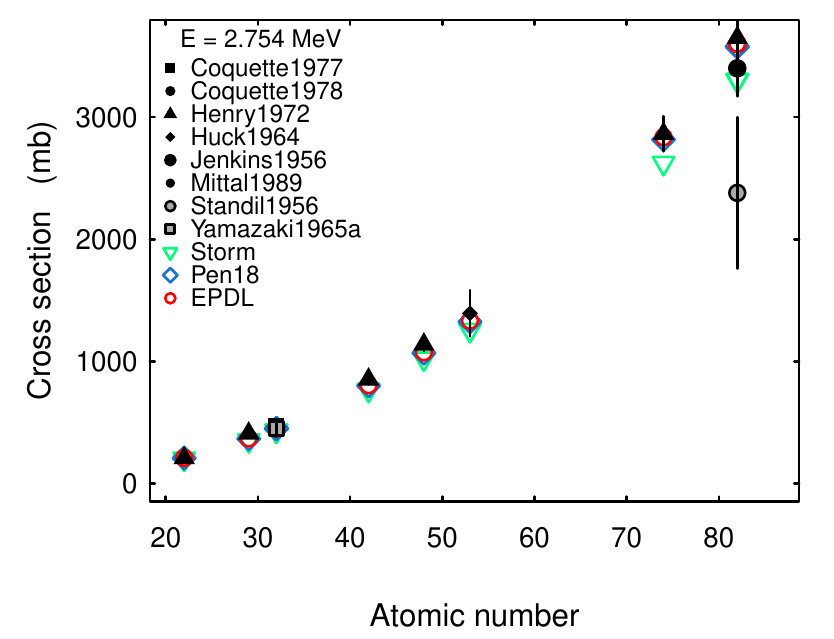}}
\caption{Total $e^+e^-$ pair production cross sections  at 2.754 MeV as a function of the atomic number Z.}
\label{fig_tote2754}
\end{figure}

A selection of representative experimental and calculated cross sections is 
shown in Figs. \ref{fig_tot13}-\ref{fig_tote2754}.
The plots allow a qualitative appraisal of 
the data; the results of the validation process derive from statistical
inference and are documented in the following subsections.


\subsection{Comparison of calculated and experimental cross sections}
\label{sec:chi2pair}

Test cases to compare cross section calculations with experimental measurements
are defined as described in Section \ref{sec:analysis}.
Since the experiments reporting cross section measurements are generally unable
to distinguish pair and triplet production events, the comparison with
experimental data concerns the sum of  pair and triplet
production cross sections.

A few criteria were applied to reinforce the quality of the experimental data
involved in the $\chi^{2}$ test: the experimental sample includes only measurements
reported in digital form in the respective publications; it excludes
experimental values expressed in terms of Bethe-Heitler cross section or other theoretical references, unless
the authors' theoretical calculations were explicitly documented, so that the
reported values could be unambiguously converted into absolute cross section
measurements.

\begin{table}[htbp]
  \centering
  \caption{Results of the $\chi^{2}$ test of $e^+e^-$ pair production cross sections}
    \begin{tabular}{lccc}
\toprule
    Calculation Source & Pass  & Fail  & Efficiency \\
\midrule
    EPDL 97 & 38    & 7     & 0.84 $\pm $ 0.05  \\
    EPICS 2017 (ENDF) & 37    & 8     & 0.82 $\pm $ 0.06 \\
    EPICS 2017 (ENDL) & 37    & 8     & 0.82 $\pm $ 0.06\\
    ENDFB VIII.0 & 37    & 8     & 0.82 $\pm $ 0.06 \\
    Penelope 2018 & 32    & 13    & 0.71 $\pm $ 0.07 \\
    XCOM  & 21    & 24    & 0.47 $\pm $ 0.07 \\
    Storm and Israel& 5     & 28    & 0.15 $\pm $ 0.06 \\
    PHOTX & 21    & 24    & 0.47 $\pm $ 0.07 \\
    GEANT 3 & 17    & 28    & 0.38 $\pm $ 0.07 \\
    Geant4 Standard & 18    & 27    & 0.40 $\pm $ 0.07 \\
\bottomrule
    \end{tabular}%
  \label{tab:chi2eff}%
\end{table}%

The outcome of the $\chi^{2}$ test comparing the calculated cross sections with
experimental measurements is summarized in Table \ref{tab:chi2eff}.
These results concern the whole range of energies of the experimental data,
with the exception of the tests concerning Storm and Israel's
compilation, which covers photon energies up to 100 MeV.

\begin{figure}[t]
\centerline{\includegraphics[angle=0,width=8.5cm]{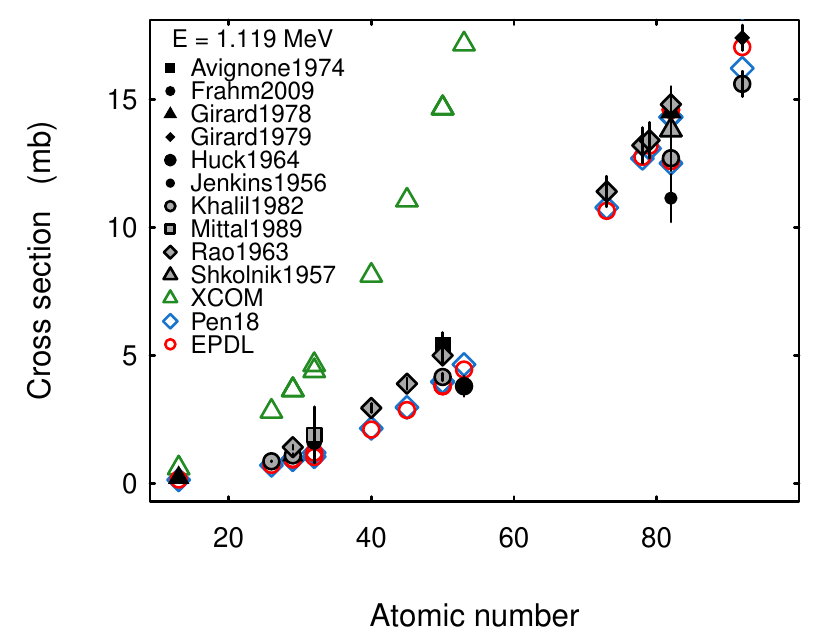}}
\caption{Total $e^+e^-$ pair production cross sections  at 1.119 MeV as a function of the atomic number Z.}
\label{fig_tote1119}
\end{figure}

Table \ref{tab:chi2effrange2} reports the results limited to experimental
measurements above 1.5 MeV, selected according to the previously 
mentioned quality criteria.
The fraction of test cases where the null hypothesis is rejected is lower in
this data sample than in the full sample: this difference could be due to better
accuracy of the calculation methods at higher energies, but also to
underestimated uncertainties in experimental measurements close to the
threshold, which are delicate and more exposed to systematic effects.
Discrepancies in the measurements at low energies  reported by different
experimental groups have been highlighted in the literature \cite{jentschel2011}; 
inconsistencies in experimental and calculated cross sections
at 1.119~MeV are qualitatively visible in Fig. \ref{fig_tote1119}.
The available information is insufficient to discern the origin of the apparently
worse compatibility with experiment at lower energies.

\begin{table}[htbp]
  \centering
  \caption{Results of the $\chi^{2}$ test of $e^+e^-$ pair production cross sections at
energies above 1.5 MeV}
    \begin{tabular}{lccc}
    \toprule
    Calculation Method & Pass  & Fail  & Efficiency   \\
    \midrule
    EPDL 97 & 33    & 2     & 0.94   $\pm $ 0.04 \\
    EPICS 2017 (ENDF) & 33    & 2     & 0.94   $\pm $ 0.04 \\
    EPICS 2017 (ENDL) & 33    & 2     & 0.94   $\pm $ 0.04 \\
    ENDFB VIII.0 & 33    & 2     & 0.94   $\pm $ 0.04 \\
    Penelope 2018 & 31    & 4     & 0.89   $\pm $ 0.06 \\
    XCOM  & 29    & 6     & 0.83   $\pm $ 0.06 \\
    Storm and Israel & 14    & 9    & 0.61   $\pm $ 0.10 \\
    PHOTX & 29    & 6     & 0.83   $\pm $ 0.06 \\
    GEANT 3 & 28    & 7     & 0.80   $\pm $ 0.07 \\
    Geant4 Standard & 28    & 7     & 0.80   $\pm $ 0.07 \\
    \bottomrule
    \end{tabular}%
  \label{tab:chi2effrange2}%
\end{table}%


\subsection{Robustness of the results}
\label{sec:chi2robust}

The effect of the quality criteria applied to the experimental sample can be
appreciated by considering the results of the $\chi^{2}$ test over a more
extensive data sample, including  experimental measurements
digitized from figures and reported in terms of non reproducible theoretical
references.
The results are reported in Table \ref{tab:chi2effdirty}, where the first row
concerns data available in digital format and unambiguously as absolute values,
identified as ``Digital, Direct''.
One can observe in Table \ref{tab:chi2effdirty} that the fraction of test cases 
where the hypothesis of compatibility with experiments is rejected is larger
when the criteria of experimental data quality are relaxed.
This outcome is reflected in a decrease of the power of the tests: for instance,
the power of the Boschloo test to correctly reject the hypothesis of equivalent
compatibility with experiment of EPDL and XCOM, with respect to the alternative
hypothesis of better performance of EPDL, drops from 0.93 to 0.79 when the
criteria of experimental data quality are relaxed.
Therefore, the above mentioned stricter quality criteria were applied to all the 
tests reported in the following.

\begin{table}[htbp]
  \centering
  \caption{Results of the $\chi^{2}$ test of $e^+e^-$ pair production cross sections,
with relaxed quality criteria of the experimental data sample}
    \begin{tabular}{lccc}
    \toprule
    Data type  & Pass  & Fail  & Efficiency   \\
    \midrule
     Digital, Direct & 38    & 7     & 0.84  $\pm $ 0.05 \\
     Digital & 43    & 14    & 0.75  $\pm $ 0.05 \\
     Digital or figure & 54    & 18    & 0.75  $\pm $ 0.05 \\
    \bottomrule
    \end{tabular}%
  \label{tab:chi2effdirty}%
\end{table}%

The concern that the varying precision of the measurements included in the
experimental sample could affect the outcome of the validation process was
addressed by evaluating the compatibility of calculated cross sections with
experimental data of different precision. 

This assessment was performed over cross section calculations based on EPDL,
1997 version.
It compared the outcome of the $\chi^{2}$ test when different portions of the
error distribution shown in Fig. \ref{fig:experr} are involved: measurements
associated with relative errors smaller or larger than the median value, and
measurements corresponding to the first and the last quartiles of the
relative error distribution.
It was qualitatively observed that the $\chi^{2}$ test results in larger
``efficiency'' when higher precision experimental measurements are involved,
i.e. with experimental errors in the first quartile and smaller than the
median; nevertheless, the hypothesis of statistically equivalent compatibility
with experiment was not rejected in any comparisons involving higher and lower
precision experimental samples, with 0.01 significance.

These assessments suggest that the conclusions of the analysis are robust.


\subsection{Comparative evaluation of calculation methods}
\label{sec:cont_pair}

The results of the $\chi^{2}$ test, documented in Table \ref{tab:chi2eff}, are fed
into contingency tables, which set the grounds for the statistical analysis
comparing the compatibility with experiment of the various categories of 
cross section calculations.

\begin{figure}
    \centering
    \begin{subfigure}[b]{0.5\columnwidth}
        \includegraphics[width=\textwidth]{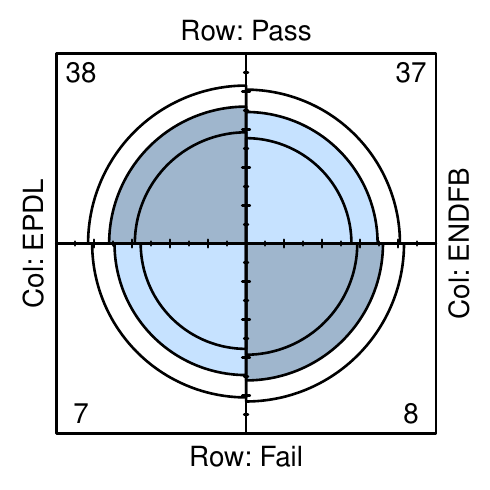}
        \caption{ENDF/B-VIII.0 and EPICS 2017}
        \label{fig_4endfb}
    \end{subfigure}%
     \begin{subfigure}[b]{0.5\columnwidth}
        \includegraphics[width=\textwidth]{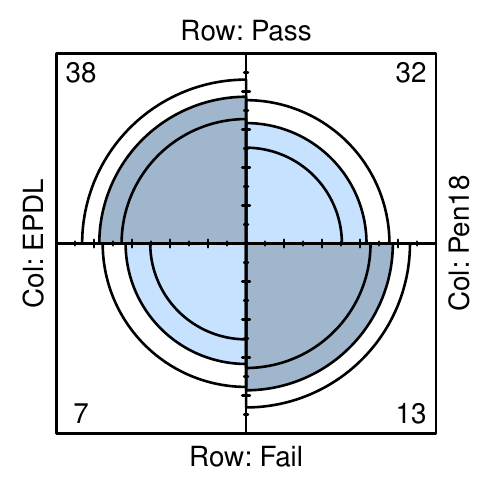}
       \caption{Penelope 2018}
        \label{fig_4Pen18}
    \end{subfigure}
    \begin{subfigure}[b]{0.5\columnwidth}
        \includegraphics[width=\textwidth]{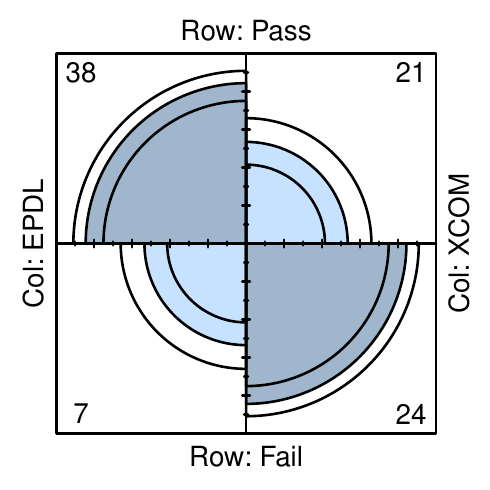}
        \caption{XCOM and PHOTX}
        \label{fig_4xcom}
    \end{subfigure}%
     \begin{subfigure}[b]{0.5\columnwidth}
        \includegraphics[width=\textwidth]{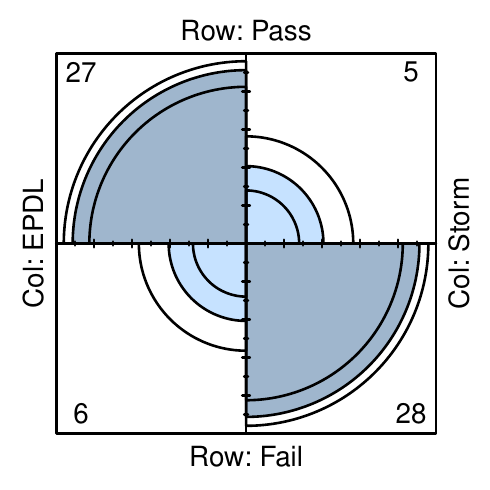}
       \caption{Storm and Israel}
        \label{fig_4storm}
    \end{subfigure}
\caption{Visual representation of 2 by 2 contingency tables summarizing the
compatibility with experiment of $e^+e^-$ pair production cross sections based
on EPDL (1997 version) and on other physics data libraries: \subref{fig_4endfb}
ENDF/B-VIII.0 and EPICS2017 in ENDF and ENDL format, \subref{fig_4Pen18}
Penelope 2018, \subref{fig_4xcom} XCOM and PHOTX, \subref{fig_4storm} Storm and
Israel.}
\label{fig_fourfold_pp}
\end{figure}

A representative selection of contingency tables is graphically
illustrated  in Fig. \ref{fig_fourfold_pp} by means of so-called ``fourfold'' plots.
The area of each quarter circle in the fourfold plots is proportional to the cell
frequency in the corresponding table; the rings represent the 99\% confidence
level for the odds ratio.
Overlapping rings in adjacent quadrants indicate consistency with the null
hypothesis that the different ``pass'' and ``fail'' counts of the $\chi^{2}$
test in the corresponding categories could arise from chance only; diagonally
opposite cells in one direction differing in size from those in the other
direction hint at qualitatively large differences in compatibility with
experiment between the two data categories.
Further documentation about fourfold plots can be found in \cite{friendly1998, 
friendly2015}; their use in a similar cross section validation context is
illustrated in \cite{tns_photoel2}.

\setlength{\tabcolsep}{3pt}
\begin{table}[htbp]
  \centering
\caption{P-values of the tests comparing the compatibility with experiment of 
different cross section calculation methods with that of EPDL97}
    \begin{tabular}{lccccc}
    \toprule
    Calculation method & Fisher & $\chi^2$ & Boschloo & Z-pooled & CSM \\
    \midrule
    EPICS 2017 (ENDF) & 1.000 & 0.777 & 1.000 & 0.860 & 0.823 \\
    EPICS 2017 (ENDL) & 1.000 & 0.777 & 1.000 & 0.860 & 0.823 \\
    ENDFB VIII.0 & 1.000 & 0.777 & 1.000 & 0.860 & 0.823 \\
    Penelope 2018 & 0.204 & 0.128 & 0.156 & 0.137 & 0.145 \\
    XCOM  & $< 0.001$ & $< 0.001$ & $< 0.001$ & $< 0.001$ & $< 0.001$ \\
    Storm and Israel & $< 0.001$ & $< 0.001$ & $< 0.001$ & $< 0.001$ & $< 0.001$ \\
    PHOTX & $< 0.001$ & $< 0.001$ & $< 0.001$ & $< 0.001$ & $< 0.001$ \\
    GEANT 3 & $< 0.001$ & $< 0.001$ & $< 0.001$ & $< 0.001$ & $< 0.001$ \\
    Geant4 Standard & $< 0.001$ & $< 0.001$ & $< 0.001$ & $< 0.001$ & $< 0.001$ \\
    \bottomrule
    \end{tabular}%
  \label{tab:cont_pp}%
\end{table}%
\setlength{\tabcolsep}{6pt}

The results of the tests applied to the contingency tables derived from the
outcome of the $\chi^{2}$ test reported in Table \ref{tab:chi2eff} are
summarized in Table \ref{tab:cont_pp}.
The tests compare the compatibility with experiment of the various cross section
calculation methods with that of the 1997 version of EPDL, which was identified
in Table \ref{tab:chi2eff} as the data library producing the highest
``efficiency''; they are consistent at rejecting the hypothesis of equivalent
compatibility with experiment for the cross sections based on XCOM, PHOTX, Storm
and Israel tabulations, and on the analytical calculations implemented in GEANT
3 and in Geant4.
The null hypothesis of equivalent compatibility with experiment is not rejected
for the cross sections based on ENDF/B-VIII.0, EPICS 2017 and Penelope 2018
tabulations.

Table \ref{tab:power_pp} documents the power of the tests to identify
significant differences in the compatibility of cross section categories with
experiment; it concerns the test cases for which the null hypothesis of
equivalent compatibility with experiment is rejected in Table \ref{tab:cont_pp}.
It reports the power of the Boschloo test, i.e. its ability to correctly reject the null hypothesis of
equivalent compatibility with experiment with respect to the two-sided
alternative hypothesis of different compatibility with experiment and to the
one-sided alternative of greater EPDL compatibility with experiment, at the
selected significance level of 0.01.
All the tests applied to contingency tables, appearing in Table
\ref{tab:cont_pp}, exhibit similar high power.

\begin{table}[htbp]
  \centering
  \caption{Power of the Boschloo test}
    \begin{tabular}{lcc}
    \toprule
    Categorical data comparison & 1-sided & 2-sided \\
    \midrule
    EPDL - XCOM/PHOTX & 0.93  & 0.77 \\
    EPDL - Storm and Israel & 1.00  & 1.00 \\
    EPDL - Geant4 Standard & 0.99  & 0.97 \\
    \bottomrule
    \end{tabular}%
  \label{tab:power_pp}%
\end{table}%


The tests over the categorical data above 1.5 MeV, deriving from the results of
the $\chi^{2}$ test reported in Table \ref{tab:chi2effrange2}, reject the null
hypothesis of compatibility with experiment equivalent to that of EPDL97 only
for the cross sections based on Storm and Israel's compilation. 
The details of the results of the tests are available in Table \ref{tab:cont_pp15}.
Also in this test case all the tests exhibit high power ($\geq$0.995).

\begin{table}[htbp]
  \centering
  \caption{-values of the tests comparing the compatibility with experiment of 
different cross section calculation methods with that of EPDL97 at
energies above 1.5 MeV}
    \begin{tabular}{lcccc}
    \toprule
    Calculation method & Fisher  & Boschloo & Z-pooled & CSM \\
    \midrule
    EPICSF & 1.000 & 1.000 & 1.000 & 0.990 \\
    EPICSL & 1.000 & 1.000 & 1.000 & 0.990 \\
    ENDFB & 1.000 &  1.000 & 1.000 & 0.990 \\
    Pen18 & 0.673 & 0.550 & 0.530 & 0.530 \\
    XCOM  & 0.259 &  0.188 & 0.152 & 0.203 \\
    Storm & $< 0.001$  &  $< 0.001$  & $< 0.001$  & $< 0.001$  \\
    PHOTX & 0.259 &  0.188 & 0.152 & 0.203 \\
    GEANT 3 & 0.151 &  0.111 & 0.086 & 0.120 \\
    Geant4 Standard & 0.151 & 0.111 & 0.086 & 0.120 \\
    \bottomrule
    \end{tabular}%
  \label{tab:cont_pp15}%
\end{table}%


One can infer from the  analysis that the ENDF/B-VIII.0, EPICS 2017
and Penelope 2018 data libraries are statistically equivalent to the 1997
version of EPDL in their ability to calculate pair production cross sections
compatible with measurements over the whole 
experimental sample.
PHOTX, XCOM (with the default energy grid), Storm and Israel's compilation, and
the analytical cross section formulations perform significantly worse than
EPDL97 when their behaviour is examined over the whole energy range covered by
the experimental data sample.
Above 1.5 MeV only Storm and Israel's compilation exhibits significantly different 
compatibility with experiment with respect to EPDL97.



\begin{figure}[htbp]
\centerline{\includegraphics[angle=0,width={0.5\columnwidth}]{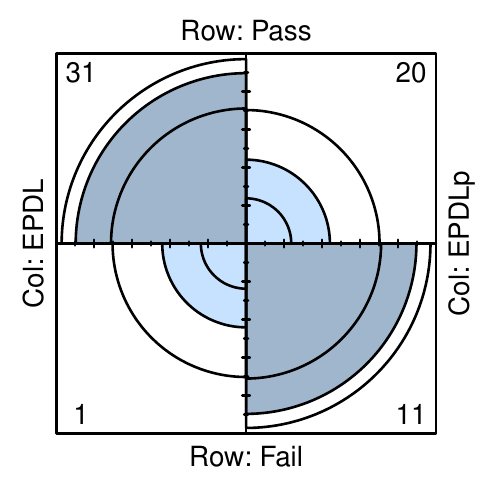}}
\caption{Visual representation of the 2 by 2 contingency table summarizing the
compatibility with experiment of EPDL97 cross sections accounting for 
 $e^+e^-$ pair production only in the field of the nucleus (EPDLp) or also accounting  for
production in the field of atomic electrons. }
\label{fig:fourtrip}
\end{figure}

\begin{figure}[htbp]
\centerline{\includegraphics[angle=0,width=8.5cm]{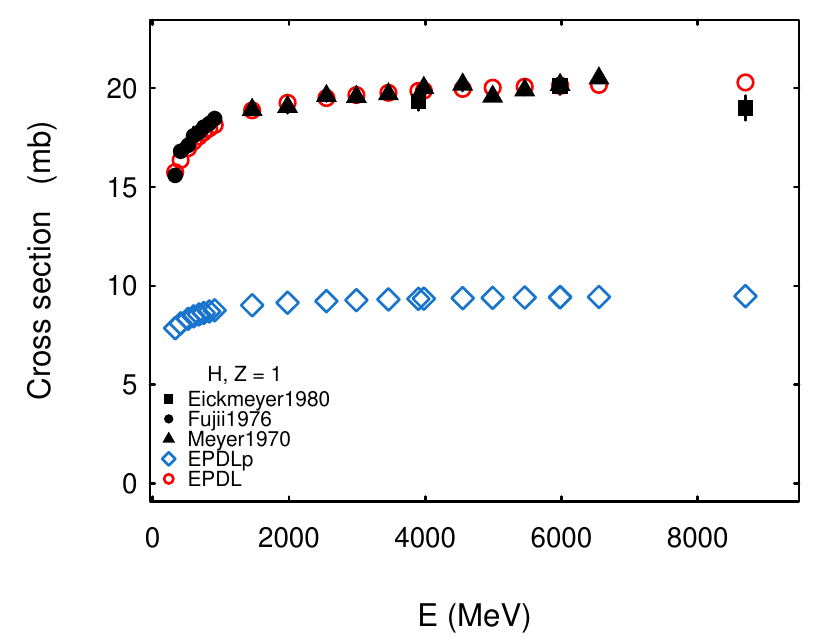}}
\caption{$e^+e^-$ pair production cross sections for hydrogen as a function of photon energy: 
EPDL calculations accounting only for pair production in the field of the nucleus (EPDLp)
and also accounting for production in the field of atomic electrons (EPDL), along with
experimental measurements.}
\label{fig:tot1_triplet}
\end{figure}

\subsection{Triplet production cross sections}

The validation of triplet production cross section calculations according to the
same methodology 
is hindered by the scarcity of precision measurements specific to this process.
Nevertheless, the validation tests provide meaningful indications regarding the
calculations of these  cross sections.

A clue comes from the $\chi^{2}$ test concerning data above the threshold for
triplet production (2.044 MeV), when one compares  experimental measurements with cross
section calculations limited to e$^+$e$^-$ pair production in the field of the
nucleus and with calculations also including the contribution of production in
the field of atomic electrons.

\begin{figure}[htbp]
\centerline{\includegraphics[angle=0,width=8.5cm]{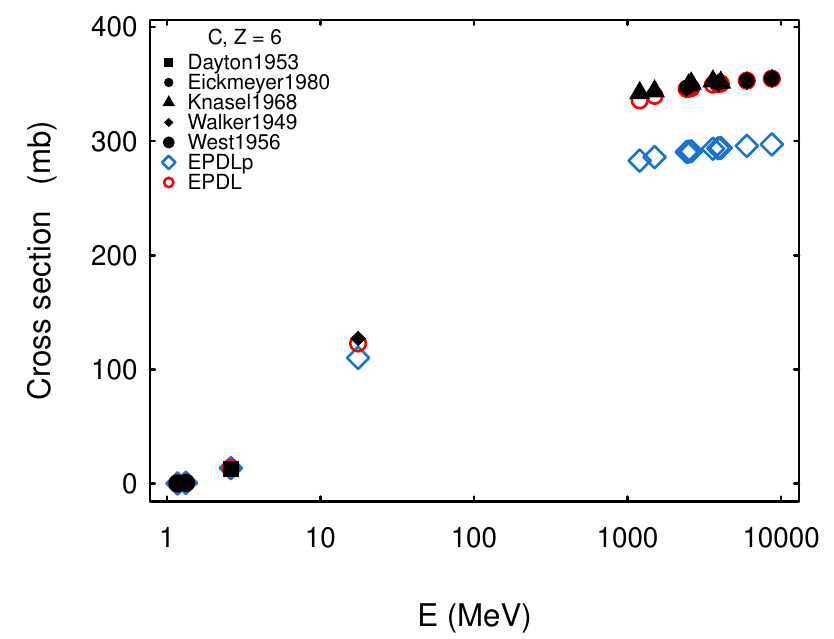}}
\caption{$e^+e^-$ pair production cross sections for carbon as a function of photon energy: 
EPDL calculations accounting only for pair production in the field of the nucleus (EPDLp)
and also accounting for production in the field of atomic electrons (EPDL), along with
experimental measurements.}
\label{fig:tot6_triplet}
\end{figure}

\begin{figure}[htbp]
\centerline{\includegraphics[angle=0,width=8.5cm]{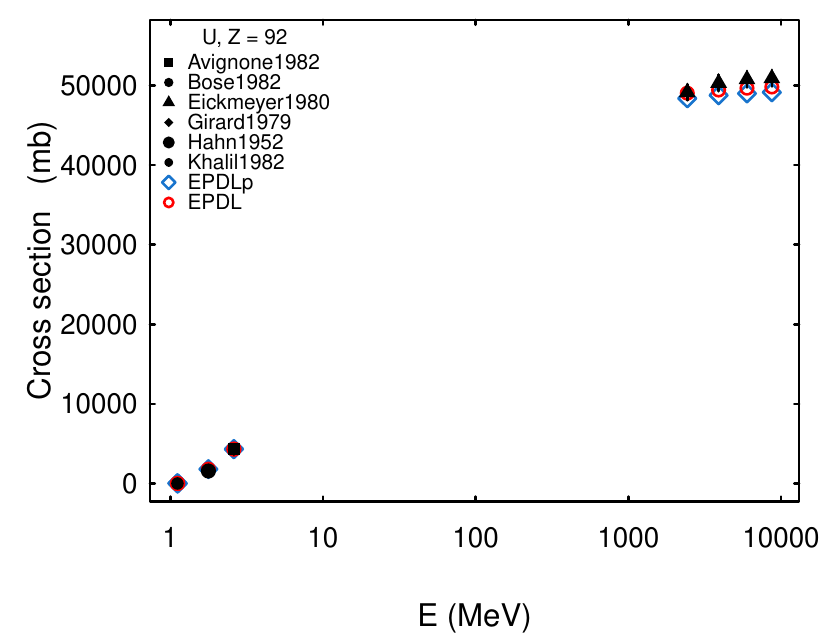}}
\caption{$e^+e^-$ pair production cross sections for uranium as a function of photon energy: 
EPDL calculations accounting only for pair production in the field of the nucleus (EPDLp)
and also accounting for production in the field of atomic electrons (EPDL), along with
experimental measurements.}
\label{fig:tot92_triplet}
\end{figure}

The results reported here concern EPDL97, but the effect is also
observed in the calculations concerning other data libraries.
The hypothesis of compatibility between experimental and calculated cross
sections is rejected in a significantly larger fraction of test cases when only
the production of e$^+$e$^-$ pairs in the field of the nucleus is accounted for:
above the threshold for triplet production, the ``efficiency'' resulting from
the $\chi^2$ test is 0.65$\pm$0.08 when accounting only for e$^+$e$^-$ pair
production in the field of the nucleus, while it is 0.97$\pm$0.03 when 
triplet production is also taken into account.
These results are illustrated in Fig. \ref{fig:fourtrip}.
The hypothesis of equivalent compatibility with experiment of the two cross
section calculations is rejected by all the exact tests applied in Section
\ref{sec:cont_pair} with 0.01 significance.

The relative contribution of pair production in the field of atomic electrons is
expected \cite{maximon1981} to be higher for elements with low atomic number,
hence, especially for photon interactions with hydrogen.
Figs. \ref{fig:tot1_triplet}, \ref{fig:tot6_triplet} and \ref{fig:tot92_triplet} confirm this prediction.

These results suggest that the contribution of triplet production in the cross
section calculation is critical to reproduce experimental measurements of
e$^+$e$^-$ pair production.


\section{Conclusion}

The data libraries and analytical parameterizations used in major Monte Carlo
codes for particle transport to calculate e$^+$e$^ -$ pair production cross
sections by photon interactions have been quantitatively evaluated regarding
their capability to reproduce experimental measurements.
Most of them originate from an authoritative compendium of theoretical
calculations \cite{hubbell_1980}; nevertheless, they exhibit some differences in
their compatibility with experiment.


The experimental data involved in the validation process derive from
an extensive survey of the literature.
Quality criteria applied to the selection of the experimental  sample,
objectively assessed by means of statistical methods, strengthen the 
reliability of the validation results.

Among the calculation methods subject to test, the statistical analysis
has identified the 1997 version of EPDL as the cross section source with the
lowest incompatibility with experiment.
More recent versions of EPDL, released in ENDF/B-VIII.0 and distributed in EPICS
2017 by IAEA, as well as the tabulations distributed with Penelope 2018, are
statistically equivalent to EPDL97 in compatibility with experiment.
EPDL97 is currently used by several Monte Carlo codes; the results of the
analysis support this choice and show that moving to more recent versions is not
needed at the present time.

Other cross section sources -- XCOM tabulations for default photon energies,
PHOTX and the parameterizations used in Geant4 Standard electromagnetic package
-- are statistically equivalent to EPDL in compatibility with experimental data
at energies above 1.5 MeV.
Their reduced ability to reproduce experimental measurements close to the 
 e$^+$e$^ -$ pair production threshold is likely due to insufficient granularity 
of the tabulations or inadequacy of the analytical parameterizations in the low
energy range corresponding to rapid variability of the cross section.
Storm and Israel's compilation, which antedates the publication of
\cite{hubbell_1980}, exhibits significant differences in compatibility with
experiment with respect to EPDL97.
Developers and maintainers of Monte Carlo codes may consider providing
alternative options of cross section calculations, if they do not already do so,
to address the shortcomings associated with some of them.
Experimental users may consider selecting appropriate cross section options in
their simulations, in general or specifically at low energies, when high
accuracy is required by sensitive applications.

The characteristics of the available experimental data and the scarcity of
specific measurements prevent the distinct appraisal of cross sections for
e$^+$e$^ -$ pair production in the field of the nucleus and in the field of
atomic electrons; nevertheless, the analysis has highlighted the significant role
played by triplet production calculations to achieve consistency with the available 
experimental data.

The scarcity of experimental data prevents stratified analyses, which could
single out the capabilities of cross section calculation methods to a finer degree.
Further measurements, also addressing the higher energy end that is currently
scarcely represented in the literature, would be helpful to refine the scope
and the depth of the validation tests.

\section*{Acknowledgment}

The authors thank Anita Hollier for proofreading the manuscript and valuable comments,
and Sergio Bertolucci for helpful support at CERN.
The CERN Library has provided substantial assistance and essential reference material for this research.

\bibliographystyle{IEEEtran}
\bibliography{IEEEabrv, biblio}

\end{document}